# Origin of Spatial Charge Inhomogeneity in Graphene

Yuanbo Zhang[1*§], Victor W. Brar[1,2*], Caglar Girit[1,2], Alex Zettl[1,2], Michael F. Crommie[1,2§]

[1]Department of Physics, University of California at Berkeley, Berkeley, California, USA

[2] Materials Sciences Division, Lawrence Berkeley Laboratory, Berkeley, California, USA

[*] These two authors contributed equally to the current work.

[§] Email: zhyb@berkeley.edu, crommie@berkeley.edu

**In an ideal graphene sheet charge carriers behave as two-dimensional (2D) Dirac fermions governed by the quantum mechanics of massless relativistic particles[1,2]. This has been confirmed by the discovery of a half-integer quantum Hall effect[3,4] in graphene flakes placed on a $SiO_2$ substrate. The Dirac fermions in graphene, however, are subject to microscopic perturbations that include topographic corrugations and electron density inhomogeneities (i.e. charge puddles). Such perturbations profoundly alter Dirac fermion behavior, with implications for their fundamental physics as well as for future graphene device applications. Here we report a new technique of Dirac point mapping that we have used to determine the origin of charge inhomogeneities in graphene. We find that fluctuations in graphene charge density are not caused by topographical corrugations, but rather by charge-donating impurities below the graphene. These impurities induce unexpected standing wave patterns due to supposedly forbidden back-scattering of Dirac fermions. Such wave patterns can be continuously modulated by electric gating. Our observations provide new insight into impurity scattering of Dirac fermions and the microscopic mechanisms limiting electronic mobility in graphene.**

Topographic corrugations and charge puddles in graphene are two of the most significant types of disorder in this new material. Topographic corrugations[5-8], for example, have been suggested as a cause for the suppression of anticipated anti-localization[9]. Electron and hole puddles[10] have similarly been blamed for obscuring universal conductivity in graphene[11]. These issues are part of a puzzle regarding the factors that limit graphene's mobility[12-18]. In order for graphene to fulfill its promise as a next generation nanodevice substrate it is important to understand the origin of this disorder and the influence it has on Dirac fermions. We have made significant progress in this direction by utilizing the techniques of scanning tunneling microscopy (STM) and spectroscopy (STS) to simultaneously probe topographic and electronic disorder in graphene with an electron density spatial resolution two orders of magnitude higher than previous measurements.

Fig. 1a displays the STM topography of a typical $30 \times 30$ nm$^2$ area on our graphene sample. We observe random corrugations with lateral dimension of a few nanometers and a vertical dimension of ~ 1.5 Å (rms), likely due to roughness in the underlying SiO$_2$ surface and/or intrinsic ripples of the graphene sheet[5-7, 19]. STM imaging at the atomic scale clearly resolves the graphene honeycomb lattice on top of the broader surface corrugation (inset).

We explored the inhomogeneous graphene charge density by spatially mapping graphene tunneling spectroscopy features. Graphene tunneling spectra exhibit a dip (local minimum) at a voltage, $V_D$, outside of a ~ 126 mV gap-like feature centered at the Fermi level for slightly doped samples (Fig. 1c, inset). $V_D$ marks the Dirac point energy, $E_D$, offset by the energy of a K point phonon mode (~ 63meV)[20]. Spatial variation in the measured value of $E_D$ reflects the spatial profile of charge inhomogeneity in graphene (two spectra taken at points separated by 17 nm, for example, are shown in the Fig. 1c inset). Charge puddles can thus be mapped by measuring the tunnel spectrum at every pixel over a given area and identifying $E_D$ at each point. A Dirac point map, $E_D(x, y)$, can be converted into a charge density map, $n(x, y)$, through the relation $n(x, y) = E_D^2(x, y) / \pi (\hbar v_F)^2$ [3, 4]. Fig. 1b displays such a map of $E_D$ for the same area shown in Fig. 1a at an applied gate voltage of $V_g = 15$ V. We clearly resolve 30 meV fluctuations in the Dirac point energy, corresponding to charge density fluctuations of $\sim 4\times 10^{11}$ cm$^{-2}$. A single puddle of electrons having a width of ~ 20 nm can be seen over this area. Integration of $n(x, y)$ over the puddle area yields a total charge inside this puddle of $0.3 \pm 0.2$ e (the average background charge density has been subtracted).

Charge puddles can also be probed by spatially mapping the quantity dI/dV for a fixed sample-tip bias held slightly below $V_D$. This technique significantly reduces data acquisition time and is particularly suited for measuring large graphene areas containing multiple charge puddles. The basis for using this second technique to measure charge puddles is illustrated in Fig. 1c. In the vicinity of the Dirac point (i.e. $V_D$) the tunneling conductance, dI/dV, is proportional to the electronic local density of states (LDOS) of graphene[20]. A spatial variation in $E_D$ therefore directly translates into a proportional spatial variation in dI/dV at a fixed bias. dI/dV maps taken at a fixed bias close to $V_D$ can thus produce a map of $E_D$, up to a multiplicative factor. This is demonstrated by the fixed-bias ($V_b = -0.25$ V) dI/dV map in Fig. 1d which shows the same charge puddle obtained from direct $E_D$ mapping (Fig. 1b). Applying this method to a larger area (topography shown in Fig. 2a), we are able to map the profile of multiple charge puddles as seen in Fig. 2b. Individual puddles having an average lateral dimension of $<L>\sim 20$ nm are clearly resolved (the electron-rich puddle outlined by a dashed black box is the same as that shown in Fig. 1b). Such puddles are prevalent in graphene, and we have used this technique to explore 23 electron-rich charge puddles over an area of 23000 nm$^2$ for three different graphene samples.

The same perturbations that create graphene charge puddles also act as scattering sites for the Dirac fermions in graphene, leading to quasiparticle interference (QPI) patterns[21-23]. This can be seen in Fig. 3a which shows a dI/dV map taken with $V_b = 0.35$ V over the same area shown in Fig. 2a. Standing wave patterns in electronic LDOS having a smaller feature size than the charge puddles are clearly resolved on top of the smooth background provided by the puddle profile shown in Fig. 2b. Dispersion in the QPI can be seen in Figs. 3b and 3c, which show the interference wavelength decrease as sample-tip bias is increased to 0.6 V and 0.85 V respectively ($V_g = 15$ V were fixed for Figs. 3a-c). We emphasize that the charge puddles are a separate phenomena from the QPI and that their size scale, $<L> \sim 20$ nm, is unrelated to the energy-dependent QPI wavelength.

These results raise two fundamental questions: (1) What specifically causes the charge puddles? and (2) how do the graphene fermions scatter from them, thus causing QPI? We now address these questions by first discussing electron scattering from the charge puddles and then by determining the actual origin of the charge puddles (we find it convenient to answer question (2) before answering question (1)).

The observed QPI patterns can be understood as the result of quasiparticle scattering from a disordered potential. This is schematically illustrated in the reciprocal space sketch of Fig. 3g where constant-energy contours cut through conical graphene bands to produce circles having energy-dependent radius $k$ around the Dirac points at K and K'. Intra-valley scattering processes caused by long-range disorder scatter the electrons across the diameter of a single constant-energy circle via a scattering wave-vector $q$ (red arrow in Fig. 3g). This results in $|q| = 2k$, i.e., electrons are back-scattered.

2D Fourier transforms of the dI/dV maps in Figs. 3a-c (shown in the insets) convert the observed spatial oscillations to reciprocal space and reveal constant-energy rings of radius $2k$.

Probing QPI as a function of $V_b$ allows us to map the 2D band structure of graphene[24]. Fig. 3d shows a radial average of the Fourier transforms in Figs. 3a-c, and it is clear that the dominant wavevector, $|q|=2k$, of the observed QPI (the radius of the ring) varies significantly as a function of $V_b$. Fig. 3e plots electron tunnel energy $E = eV_b$ versus $k = q/2$ (red dots) from such analysis and reveals a linear dispersion relation for states above ($eV_b > V_D$) and below ($eV_b < V_D$) the Dirac point ($V_g = 15$ V leads to a fixed $V_D = -0.2$ V for this measurement). Fitting this data with the expected graphene dispersion relation, $E = \hbar v_F k$, we obtain $v_F = 1.5 \pm 0.2 \times 10^6$ and $1.4 \pm 0.2 \times 10^6$ m/s for states above and below the Dirac point, respectively.

The gate-tunability of graphene also provides a unique opportunity to probe the energy dependence of the QPI *without* changing the STM sample-tip bias. QPI patterns obtained in this way for fixed $V_b = 0.75$ V and a changing $V_g$ were Fourier analyzed as above, resulting in a $k$ versus $V_g$ dispersion that is plotted in Fig. 3f. From the linear band structure of graphene (including the inelastic phonon offset, $\hbar\omega_0 = 63$ meV) we expect this gate-dependent dispersion to have the following form:

$$k = \frac{eV_b - \hbar\omega_0}{\hbar v_F} + \text{sgn}(n)\sqrt{\pi|n|}, \quad n = \alpha(V_g - V_0), \tag{1}$$

where $n$ is the net charge carrier density induced by both the gate ($V_g$) and the environment ($V_0$) assuming a simple parallel capacitor model. Here $\alpha = 7.1 \times 10^{10}$ cm$^{-2}$/V is estimated from the device geometry and $V_0 \approx 0$ V can be obtained from gate-dependent spectroscopic measurement[20]. Using the value for $v_F$ obtained from the data in Fig. 3e, we find that Eq. (1) fits our measured gate-dependent dispersion quite well with no adjustable parameters (Fig. 3f, solid red line).

Our QPI-based electronic dispersion measurement differs from theoretical expectations and previous experimental measurements of epitaxial monolayer[25] and bilayer[24] graphene grown on SiC. There are three main points to notice. First, a gap exists in our experimental dispersion relation (Fig. 3e) at $k = 0$ which we attribute to energy loss to the $\hbar\omega_0 = 63$ meV phonon modes during inelastic electron tunneling[20]. Second, our extracted band slopes are $\sim 32 \pm 18\%$ bigger than what we expect from the commonly accepted value of the graphene Fermi velocity, $v_F = 1.1 \times 10^6$ m/s[3, 4] (this "theoretical" value of $v_F$ leads to the poorly fitting dashed blue lines in Figs. 3e and 3f, with the inelastic phonon energy loss taken into account). We note that angle-resolved photoemission spectroscopy measurements of graphene also result in a similarly increased slope in the filled states, which has been attributed to band renormalization due to plasmons[26].

The third intriguing aspect of our observed QPI is the fact that we see backscattering at all. Theoretical models that take Dirac fermion pseudo-spin into account suggest that intra-valley backscattering processes are forbidden in monolayer graphene[2, 27] (in sharp contrast to bilayer graphene where intra-valley backscattering processes are allowed[24]). Intra-valley backscattering was recently reported as absent in monolayer graphene epitaxially grown on SiC, and pesudo-spin-suppressed backscattering was provided as an explanation[25]. In general, some intra-valley backscattering is expected to occur as a second order process that is predicted to lead to a fast decay of standing wave patterns[28] ($\sim 1/r^2$, where $r$ is the distance from the scatterer). Our observations of intra-valley backscattering, however, do not clearly support such fast decay behavior, suggesting that there may be other symmetry-breaking mechanisms at work in graphene flake samples[29].

We are now poised to explain the origin of the charge puddles, which is also the origin of the scattering-induced QPI that we observe. We first rule out the hypothesis that topographic corrugations in graphene are a primary cause of the charge puddles. A comparison between the geometry of the charge puddles we observe (Fig. 2b) and topographic corrugations over the same area (Fig. 2a) yields no apparent correlation, as the puddles are an order of magnitude larger than the size of the topographic corrugations. We have also computed the curvature of the graphene monolayer characterized by the Laplacian of the topography, $\nabla^2 z(x,y)$. Fig. 2c shows a map of the curvature over the same surface area as Fig. 2a. The average feature size in the curvature map is more than an order of magnitude smaller than that of the charge puddles, further ruling out surface corrugation as the cause of the puddles.

There is, however, a strong correlation between highly localized features seen in our large bias dI/dV maps and the charge puddles. These localized scattering centers show up as "dots" in the QPI patterns and occur only in electron-rich charge puddles when the electron wavelength is reduced by large bias, as shown by the red arrows in Fig. 3c. We have observed such localized scattering centers in all of the electron-rich puddles that we have tested. For example, in Fig. 4a we show STM topography of a different region of the graphene surface that exhibits typical charge puddles in dI/dV maps obtained at sample-tip biases very close to the Dirac point (see Fig. 4b). When the bias is moved away from the Dirac point, as shown in Fig.4c, we clearly see local scattering centers in the electron-rich regions of these charge puddles. Because the scattering centers do not correspond to any clear topographical features, we believe that they arise from individual charged impurities located beneath the graphene. This interpretation is supported by recent experiments on suspended graphene sheets[30, 31].

In order to gain deeper insight into the origin of these subsurface impurities, we performed numerical integration of the charge in five different charge puddles and compared the total amount of charge per puddle to the number of impurities per puddle. In Fig. 4d we plot the total charge of the puddles as a function of the number of impurities they contain. This data falls roughly on a line, the slope of which allows us to estimate that the average charge contributed by an individual impurity is $\sim 0.07 \pm 0.03$ e[32]. Interestingly, some calculations[33, 34] predict a charge transfer of this order when molecules from air (such as $N_2$ and $H_2O$) are physisorbed onto graphene. This finding, combined with the fact that our samples are prepared in ambient conditions, provides further evidence that molecules from air trapped between graphene and the $SiO_2$ substrate are the likely origin of the charge puddles that we observe in graphene flake nanodevices.

In conclusion, we have imaged the nm-scale charge landscape that Dirac fermions experience as they move through graphene. We show directly that charge puddles having an average lengthscale of 20nm arise from charge-donating impurities. Electronic scattering from these charge fluctuations leads to unexpected back-scattering processes that appear to violate pseudo-spin conservation. These findings give us new insight into the microscopic processes that limit electron mobility in graphene flakes, and point toward new strategies for improving graphene nanodevice behavior.

**Methods**

Our graphene monolayer flakes were prepared on an oxidized Si wafer in a similar fashion as described in Ref.[35]. We made electrical contact to graphene by direct deposition of 12 nm thick Ti (or Au) electrodes through a stencil mask to avoid photoresist contamination. Heavily doped Si under a 285nm $SiO_2$ layer was used as a back-gate, allowing us to vary the carrier density in the graphene sample. As part of a cleaning procedure the samples were annealed at 180 ºC in ultra-high vacuum (background pressure < $10^{-10}$ mbar) for ~ 10 hours. *In situ* electric transport measurements have shown that the graphene samples prepared in this way have a typical mobility of ~ 6000 $cm^2$/Vs (see the Supplementary Information).

Experiments were conducted with a modified Omicron LT-STM at low temperature (T = 4.8 K) and in a UHV environment with base pressure < $10^{-11}$ mbar. We find that the preparation of STM tips is crucial for reliable spectroscopic measurement on graphene. To ensure that our STM tips were free of anomalies in their electronic structure, we calibrated the tips by performing tunneling differential conductance (dI/dV) measurements on a clean Au(111) surface both before and after graphene measurement. dI/dV spectra were measured using lock-in detection of the AC tunnel current, *I* , after adding an 8 meV (rms) modulation at 517 Hz to the sample bias voltage $V_b$.

**Acknowledgements**

We thank D.-H. Lee, H. Zhai, S. Louie, A. Balatsky, J. Moore, A. Vishwanath, F. Wang, and C. H. Park for helpful discussions. This work was supported by DOE under contract No. DE-AC03-76SF0098. Y.Z. acknowledges a postdoctoral fellowship from the Miller Institute, UC Berkeley.

**Figure Captions:**

**Figure 1. STM topography and charge puddle profile of graphene. a,** STM topograph ($V_b = -0.25$ V, $I = 20$ pA) of a $30 \times 30$ nm$^2$ patch of graphene resting on a SiO$_2$ substrate. Inset: Close-up topograph of the graphene honeycomb lattice. **b,** Dirac point energy ($E_D$) map of a single charge puddle lying in the same area shown in **a** ($V_g = 15$ V). This is converted to a local charge density map of graphene (an average background charge density of $0.9 \times 10^{12}$ cm$^{-2}$ has been subtracted). **c,** Sketch showing how changes in the Dirac point energy ($\Delta E_D$) are proportional to changes in dI/dV signal intensity ($\Delta$dI/dV) at a fixed sample-tip bias. Inset: dI/dV spectra taken at two points separated by 17 nm on a graphene surface. Positional change in Dirac point energy can be seen. **d,** Fixed bias dI/dV map over same area as **a** and **b** shows same puddle profile for same V$_g$..

**Figure 2. Large area image of graphene topography and charge puddles. a,** $60 \times 60$ nm$^2$ constant current STM topograph of graphene ($V_b = -0.225$ V, $I = 20$ pA). **b,** dI/dV map ($V_b = -0.225$ V, $I = 20$ pA, $V_g = 15$ V) taken simultaneously with **a** reveals electron puddles with a characteristic length of ~ 20 nm. **c,** Curvature of surface obtained by calculating the Laplacian of the topographic image shown in **a**. Upper left dashed boxes indicate the same area shown in Fig. 1.

**Figure 3. Quasiparticle scattering on a graphene surface. a-c,** dI/dV maps of the same area shown in Fig. 2 obtained at $V_b = 0.35$, 0.6 and 0.85 V respectively. The tunnel

current is held at $I = 50$, 60 and 70 pA respectively and the gate voltage is fixed at $V_g = 15$ V for all three measurements. Lower right insets: 2D Fourier transform of each image. Upper left dashed boxes indicate the same area shown in Fig. 1. Red arrows in **c** point to localized scattering centers. **d,** Radial averaged intensity profiles of the 2D Fourier transforms shown in **a-c** plotted as a function of $k$. Red lines indicate Lorentzian fits. Curves are vertically displaced for clarity. **e,** Quasiparticle energy dispersion above and below the Dirac point ($V_D = -0.2$ V, $V_g = 15$ V). Each point is extracted from a Fourier analysis as in **a-d**. Solid red lines show fitted linear curves yielding $v_F = 1.5 \pm 0.2$ and $1.4 \pm 0.2 \times 10^6$ m/s for upper and lower branches. Blue dashed lines indicate theoretical dispersion for $v_F = 1.1 \times 10^6$ m/s [3,4] assuming 63 meV offsets due to phonon assisted inelastic tunneling[20] **f,** Gate dependence of $k$ as a function of $V_g$ at a constant sample-tip bias of $V_b = -0.75$ V (each point is extracted from a Fourier analyzed dI/dV map). Solid red line shows the calculated dispersion obtained using eq. (1) and $v_F$ as measured in **e**. Dashed blue line shows the theoretical dispersion arising when $v_F = 1.1 \times 10^6$ m/s. **g,** Schematic of the 2D Brillouin zone of graphene with orange circles indicating constant energy contours for states around the K and K' points near the Fermi energy. The scattering wavevector for an intra-valley back-scattering process is given by $q$.

**Figure 4. Impurities in Graphene. a,** STM topography of $50 \times 50$ nm$^2$ area of graphene. **b,** dI/dV map at bias near Dirac point ($V_b = -0.29$ V, $I = 25$ pA, $V_g = 15$ V) shows electron puddles due to charge fluctuations over same region of graphene as **a**. Red

crosses indicate the location of quasiparticle scattering center impurities observed in **c**. **c,** dI/dV map of same area at larger bias ($V_b = -0.75$ V, $I = 80$ pA and $V_g = 60$ V) reveals impurity scattering centers in electron-rich charge density puddles (red crosses). **d,** Integrated charge per electron puddle plotted as a function of the number of observed impurities in each puddle (puddles are defined as the electron-rich regions left after subtracting the average background charge density). Linear fit to the data (black line) gives charge per impurity as $0.07 \pm 0.03$ e ("e" is the charge of an electron).

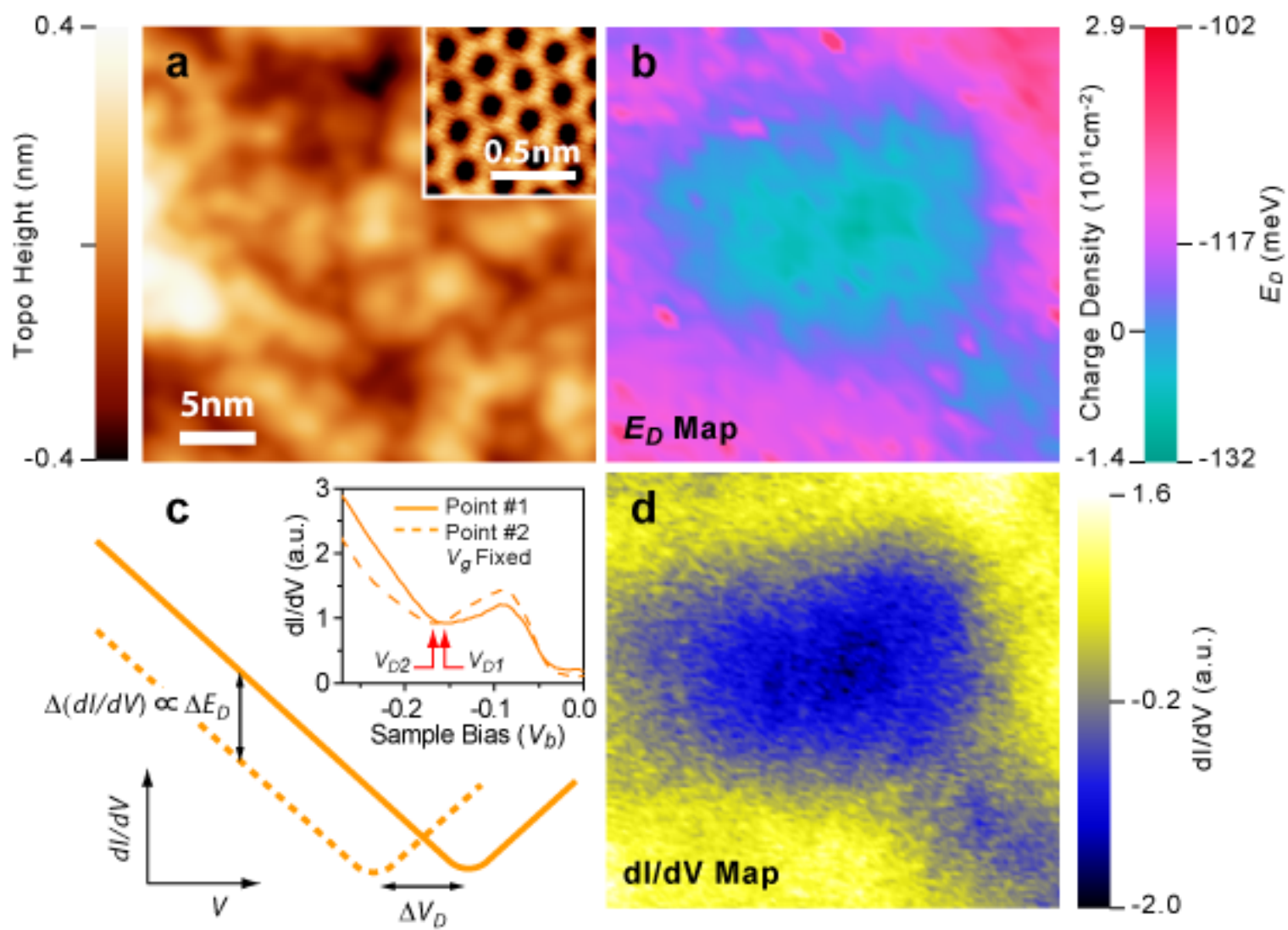

Figure 1

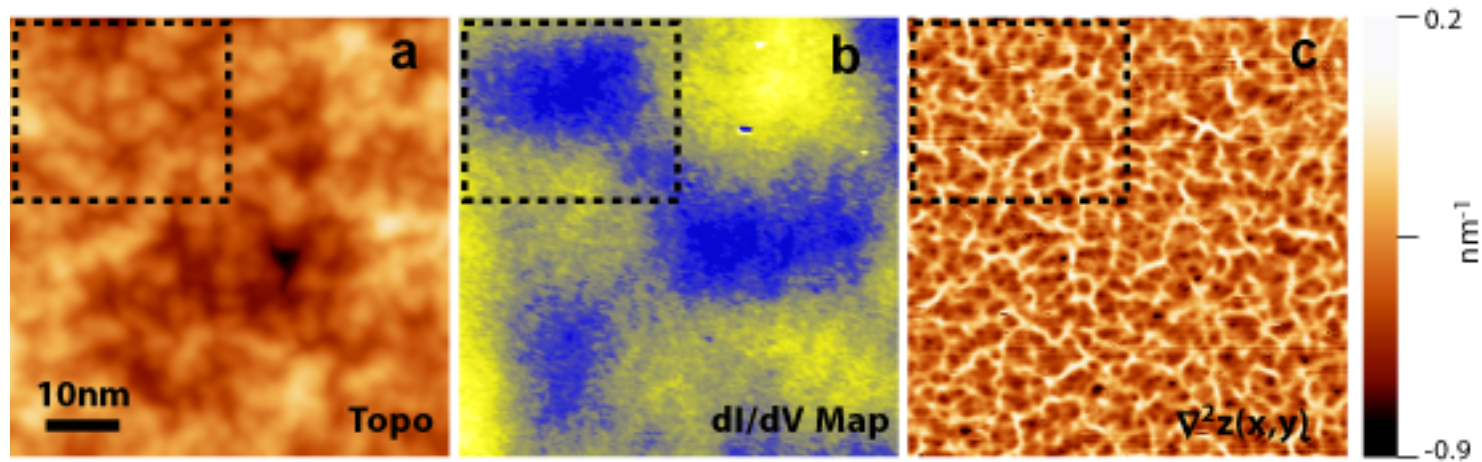

Figure 2

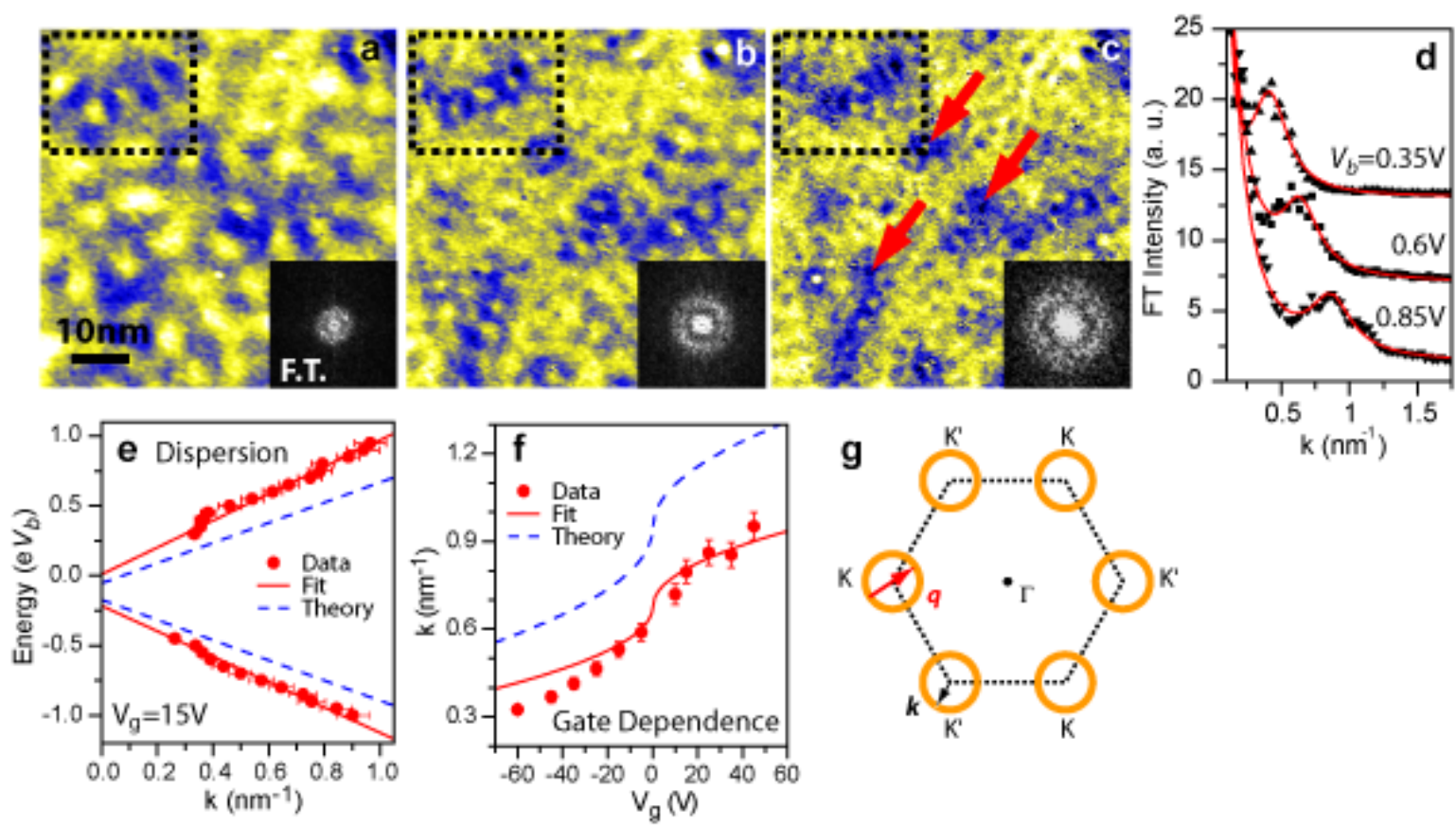

Figure 3

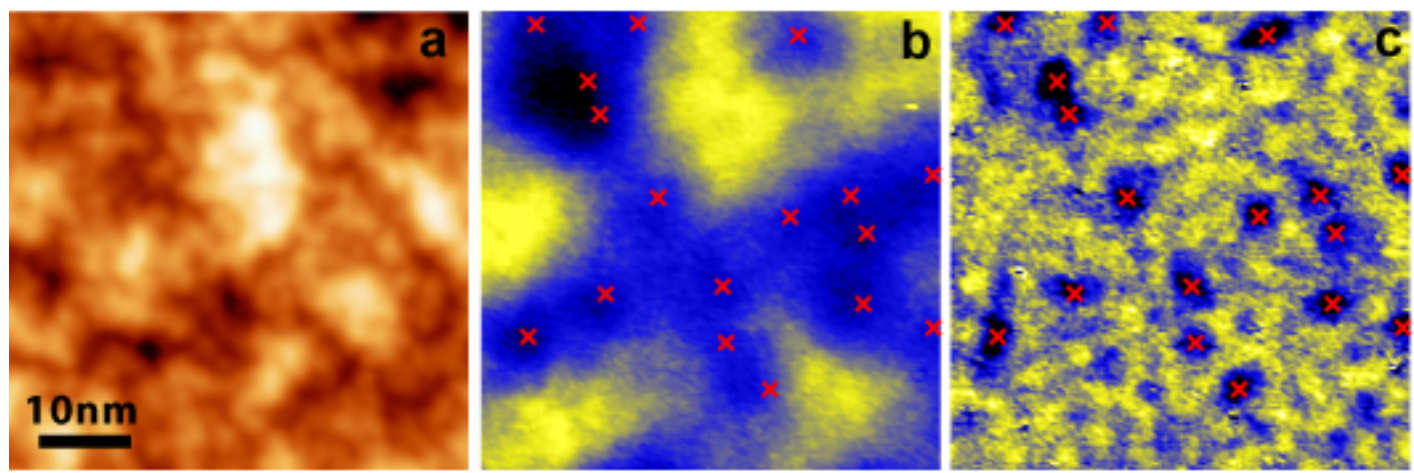

Figure 4